\documentclass[letterpaper]{article} 
\usepackage{aaai23}  
\usepackage{times}  
\usepackage{helvet}  
\usepackage{courier}  
\usepackage[hyphens]{url}  
\usepackage{graphicx} 
\usepackage{natbib}  
\usepackage{caption} 
\usepackage{algorithm}
\usepackage{algorithmic}
\usepackage{graphicx}
\usepackage{tabularx}

\usepackage{newfloat}
\usepackage{listings}
\DeclareCaptionStyle{ruled}{labelfont=normalfont,labelsep=colon,strut=off} 
\floatstyle{ruled}
\newfloat{listing}{tb}{lst}{}
\floatname{listing}{Listing}

\title{\centering\bfseries\fontsize{14}{16.8}\selectfont AI-driven E-Liability Knowledge Graphs: A Comprehensive Framework for Supply Chain Carbon Accounting and Emissions Liability Management}

\author{
    Olamide Oladeji\textsuperscript{\rm 1,2}, Seyed Shahabeddin Mousavi\textsuperscript{\rm 1,2}, Marc Roston\textsuperscript{\rm 2}\\\
}
\affiliations{
    \textsuperscript{\rm 1}Department of Management Science \& Engineering, Stanford University, Stanford, CA, USA\\
    \textsuperscript{\rm 2}Sustainable Finance Initiative, Precourt Institute for Energy, Stanford University, Stanford, CA, USA\\

    oladeji@stanford.edu, ssmousav@cs.stanford.edu, marc.roston@stanford.edu
}

\usepackage{bibentry}

\begin{document}

\maketitle
\begin{abstract}
\begin{quote}

While carbon accounting plays a fundamental role in our fight against climate change, it is not without its challenges. We begin the paper with a critique of the conventional carbon accounting practices, after which we proceed to introduce the E-liability carbon accounting methodology  and Emissions Liability Management (ELM) originally proposed by Kaplan and Ramanna  in \cite{kaplan2021accounting}, highlighting their strengths. Recognizing the immense value of this novel approach for real-world carbon accounting improvement, we introduce a novel data-driven integrative framework that leverages AI and computation - the E-Liability Knowledge Graph framework - to achieve real-world implementation of the E-liability carbon accounting methodology. In addition to providing a path-to-implementation, our proposed framework brings clarity to the complex environmental interactions within supply chains, thus enabling better informed and more responsible decision-making.
We analyze the implementation aspects of this framework and conclude with a discourse on the role of this AI-aided knowledge graph in ensuring the transparency and decarbonization of global supply chains.
 
\end{quote}
\end{abstract}

\section{Introduction}

Robust carbon accounting is very vital to the goal of cutting carbon emissions by providing a clearer picture of the sources and magnitude of emissions. Among other uses, rigorous carbon accounting can enable us to identify the most effective areas for climate change mitigation and also aid in the monitoring of progress of global mitigation efforts. Despite the increasing adoption of various carbon accounting methodology such as the GHG Protocol's Corporate Accounting Standards, by organizations and governments around the world, a number of issues have necessitated that we rethink modern carbon accounting methodology in order to realize its true potential. 
We highlight below, some of the limitations of current carbon accounting practices:

\begin{itemize}
    \item \textbf{Complexity of emissions sources:} Carbon accounting can be very complex due to the vast range of emission sources that exit. Such emitting sources can include everything from large-scale industrial processes to the burning of fossil fuels for electricity and transportation, deforestation, and even natural events such wildfires. With each source having distinct characteristics and requiring different methodologies to accurately measure its emissions, robustly accounting for emissions may sometimes be a daunting challenge. Another aspect that affects this is the interwoven nature of most supply chains\cite{jiang2019factors}. Manufacturing a single product or providing a service will typically require complex interactions and exchange of goods and services, all guided by complex technological processes. Robust carbon accounting methodology must sufficiently capture these complexities however they present in the supply chain. 

    \item \textbf{Disparate, unavailable data:} Another major limitation associated with current carbon accounting practices is that the data necessary for carbon accounting is often scattered across various repositories, or in some cases, not available. This necessitates significant efforts to gather, verify, and standardize the data. Beyond data gathering, data often needs to be updated regularly to accurately reflect evolving realities. Several research studies have described this data limitation challenge. For example, the authors of \cite{ducoulombier2021understanding} review the data limitations associated with carbon accounting in significant detail.

    \item \textbf{High uncertainty:} One of the biggest challenges is the inherent uncertainty in the measurement of greenhouse gas emissions. It is often not possible to directly measure emissions, especially at large scales. Instead, researchers, organizations and policymakers often have to rely on estimates and indirect measurements which are subject to varying degrees of uncertainty. Such uncertainty is an inherent part of carbon accounting and needs to be adequately managed to ensure the reliability of the results. Examining the literature, we can see that many criticisms have been made on current practices not properly addressing uncertainty issues \cite{brander2021carbon}. For example, Williamson et. al in \cite{williamson2022carbon} criticize the integration of uncertainty in carbon accounting for coastal ecosystems.  \cite{he2022corporate} also presents a systematic review of corporate carbon accounting research, highlighting insufficient management of uncertainty as a significant issue. 

    \item \textbf{Limited, duplicative counting methodology:} The most popular framework for accounting and reporting carbon emissions at the organizational level is the GreenHouse Gas Protocol Corporate Standard. The Greenhouse Gas Protocol (GHG Protocol) was established as an initiative of the World Resource Institute and the World Business Council for Sustainable Development (WBCSD) in 1998. It published its first version of Corporate Standards categorizing corporate emissions reporting  into Scope 1, Scope 2, and Scope 3, setting the stage for what has become the de facto standard for carbon accounting at the corporate and national level\cite{wbcsd2004greenhouse}. While it has encouraged carbon emissions reporting and provided direction where there was none, the delineation into Scope 1-2-3 emissions by the GHG protocol has inadvertently led duplicative counting of emissions, among other issues \cite{williamson2022carbon}. 
\end{itemize}

In an attempt to remedy these and many other issues, the concept of E-liability based Carbon Accounting was introduced by Kaplan and Ramanna in \cite{kaplan2021accounting}. In \cite{kaplan2021accounting}, the authors propose the E-liability accounting system to address the duplicative counting and transparency issues associated with the existing GHG Protocol Scope 1-2-3 carbon accounting framework. Under this framework, where emissions are measured using a combination of chemistry
and engineering, and the principles of cost accounting are
applied to assign the emissions to individual outputs. The
authors provide a detailed method for assigning E-liabilities across an entire value chain, using the example of a car-door manufacturer whose furthest-removed supplier is a mining company, which transfers its products to a shipping company, which transports them to a steel company, and so on until the car reaches the end customer. 
The E-liability accounting allocates carbon emissions in an approach that is similar to how cost accounting distributes different expenses across the stages of production. It does this by tracking the costs in terms of carbon emissions, quantified in tons of CO2 thus serving as a way to quantify the `carbon cost' of a product or service.
Under this model, all inputs into a product, whether raw materials or intermediary goods, are deemed to carry an E-liability. This e-liability refers to the estimated amount of carbon emissions associated with that input's production and supply. As a firm goes about its production process, it adds these inherited E-liabilities to its own direct (Scope 1) emissions. Thus, the inherited E-liabilities and the firm's own direct emissions together give a more complete picture of the carbon cost of production for the firm for that product.
Having described E-liabilities, we now discuss how they might be implemented in real-world supply chains. 

\section{Leveraging Knowledge Graphs and AI for E-Liability Carbon Accounting in Supply Chains}

\subsection{Introduction to Knowledge Graphs}
Knowledge graphs (KGs) are emerging as a powerful tool for integrating, organizing, and analyzing complex, multi-relational data. First popularized by Google's Knowledge Graph which extracts and integrates information from a variety of sources to enhance its search results \cite{fensel2020introduction}, they have since been adopted in diverse fields such as bioinformatics \cite{mohamed2020discovering}, e-commerce \cite{xu2020product}, and social network analysis \cite{qian2017social}.  Fundamentally, a knowledge graph can be defined as a graph-based data model where nodes represent entities and edges denote the relationships or properties between these entities. Such a data model facilitates the encoding of rich semantic relationships between diverse entities and thus going beyond the capabilities of traditional relational databases. In addition to these capabilities, knowledge graphs can also capture the attributes of these entities and relationships, leading to a highly detailed and context-aware data representation \cite{fensel2020introduction}. One of the key advantages of knowledge graphs is their ability to handle heterogeneous and multi-modal data, which is often the case in real-world scenarios. They can accommodate various data types and structures, providing much-needed flexibility in data representation and integration. This flexibility, coupled with their semantic richness makes them well-suited for complex analytical tasks, such as anomaly detection, entity resolution, among others \cite{fensel2020introduction}.

Despite these advantages, the development and maintenance of KGs are not without challenges. For example, data quality must be ensured when working with these graphs, as is the management of uncertainty. In addition, considering the fact that knowledge graphs must enable efficient querying and manipulation, while preserving privacy and security, this can be a complicated effort. However, with the rapid advancements in graph databases and machine learning techniques, many of these challenges are being actively addressed \cite{peng2023knowledge}. 

Having introduced these graphs, we now discuss their potential role in the context of E-liability tracking and emissions liability management. 

\subsection{The Need for Knowledge Graphs in Carbon Accounting}
A crucial challenge in carbon accounting, especially in the context of  E-liabilities and ELM, is the inherent complexity and heterogeneity of the data involved. Emissions data span multiple dimensions – they can be associated with various activities (such as manufacturing or transportation), be tied to specific products or services, and occur at different stages of the supply chain. Moreover, the data come from diverse sources, which can range from company reports and governmental databases to sensor readings and satellite images \cite{wiedmann2018environmental}. As we have previously mentioned, currently existing accounting protocols, such as those disparately counting Scope 1, 2, and 3 emissions, often lead to duplicative counting or misrepresentation of total emissions due to their limitations in handling this complex data landscape\cite{MONYEI201848}\cite{brander2018creative}. For example, as Brander et al. point out in \cite{brander2018creative}, some current market-based practices whereby an organization purchases contractually, the right to claim emissions factors associated with a renewable project, and then uses that for its Scope 2 generation reporting purposes is very misleading.
We have previously introduced the E-liabilities carbon accounting methodology proposed by Robert Kaplan and Karthik Ramanna as an effective approach to mitigate many of these issues. However, from an implementation standpoint, such a method will require the integration and management of heterogenous data types from various sources and a robust computational framework to track and evaluate a firm's current E-liabilities. This need for a robust and flexible data framework becomes paramount due to the multifaceted nature of supply chains and the dynamic aspects of E-liabilities. The absence of such a framework could mean that the emissions liability management proposals such as those by Roston et. al in \cite{roston2022road} remain theoretical and academic exercises. The heterogeneity of supply chain data —spanning from raw materials to finished products, involving multiple entities such as suppliers, manufacturers, distributors and end consumers, and containing diverse data types like quantities, costs, emissions, and more— demands an associated data model that can capture these complexities. 

This is where knowledge graphs shine. Knowledge graphs, given their ability to handle multi-relational and multi-modal data, are well suited to meet these requirements. Their graph-based structure allows for the representation of the intricate dependencies between different entities in the supply chain and the various dimensions of emissions data.  As we have previously mentioned, they provide a highly adaptable and semantically rich representation of data, accommodating the multi-dimensional attributes of the supply chain. In addition, knowledge graphs can also support a more granular and accurate tracking of E-liabilities across supply chain processes because they are designed to handle the `inheritance' and accumulation of E-liabilities from suppliers to products and to buyers, offering a holistic view of a product's carbon footprint. Lastly, the semantic richness of KGs enables capturing the nuances and contexts in emissions data, thereby enabling a more accurate and comprehensive carbon accounting \cite{heath2011linked}.
\\
The rich semantics of knowledge graphs also have the potential to improve decision-making processes for Emissions Liability Management. Under an E-liability accounting framework, various stakeholders need to make important decisions to manage emissions and mitigate climate change. Beyond facilitating how each firm can easily and transparently compute its emission, E-liability based knowledge graphs can also help with policymaking and climate mitigation research efforts. For example, policymakers may need to introduce new regulations informed by learning from E-liability based carbon accounting practices. An E-liability knowledge graph of an industry could facilitate insights on how E-liabilities are distributed across different parts of the supply chain, which suppliers or products have the highest E-liabilities, and how changes in one part of the supply chain might impact the overall E-liabilities of firms in the industry. There are other potential uses of any such knowledge graph which is able to capture the emissions and interactions at a detailed yet large-enough scale. Some of these other uses are enumerated below:

\begin{itemize}
    \item Enabling independent verification of carbon emissions.
    \item Facilitating climate and macroeconomic research and technology development.
    \item Informing standard-setting.
\end{itemize}

Overall, we can see that the potential of knowledge graphs in ELM is considerable and its integration for real-world decision-making warrants further exploration. However, developing a KG-based carbon accounting framework requires careful consideration of several aspects. These include identifying suitable data sources, developing effective methods for data extraction and integration, defining the structure and semantics of the KG, and ensuring data quality, security, and privacy. In subsequent subsections, we discuss these aspects in detail and propose a roadmap for developing a KG-based carbon accounting framework.

\subsection{Designing the AI-driven E-Liability Knowledge Graph}
Here, we propose an approach to design the conceptual E-liability Knowledge Graph. Designing a knowledge graph to support emissions liability management involves several considerations. We need to decide what entities and relationships to represent, as well as what data to store about each entity and relationship. The core this graph would be entities that play a part in the lifecycle of products and services, namely, the organizations, products, and services themselves. Each organization entity would represent a unique company or institution, storing data about its location, industry, size, and carbon emissions data if available. On the other hand, product entities would represent distinct goods or services. These entities would carry data about the product's life cycle emissions which would encompass Scope 1, Scope 2, and Scope 3 emissions. Other relevant product attributes such as product category, raw material inputs, and information about the production process could also be included. This is crucial since different products have vastly different carbon footprints even within the same company.
Service entities would store similar data to products but with customization that allows us to cater for services rather than physical products. For example, they would need to include data about the nature of the service, and its associated carbon footprint derived from the energy used to deliver the service and the resources utilized in the process.

The edges in this graph are relationships and they would encode the transactions between organizations (such as sales and purchases), the production of products by organizations, and the provision of services. These relationships are very important as they allow us to trace the flow of products and services along the supply chain, while also capturing the transfer of E-liabilities. The data stored about each transaction might include the quantity of the product or service transacted, the associated carbon footprint, as well as the timing of the transaction.

The design of this knowledge graph is dictated by the need to manage the aggregated E-liabilities via Emissions Liability Management practices such as those discussed in \cite{roston2022road} and \cite{kaplan2023accounting}. To accurately track E-liabilities, we need comprehensive data about the carbon footprints of all products and services, and we need to trace the flow of these goods and services through the economy. By representing organizations, products, services, and transactions as entities and relationships in a knowledge graph, we can create a flexible, scalable model that accommodates this complexity. Furthermore, the semantic nature of knowledge graphs enables capturing nuances in relationships, critical for accurate carbon accounting. Altogether, these then inform emissions liability management decisions such as the investments in removal offsets. 

It is very important to mention that the design of such a knowledge graph should be iterative. As we learn more about the carbon emissions associated with different industries, products, and services, we would need to refine the data stored about each entity and relationship. Similarly, as we understand more about how carbon emissions are transferred through the economy, we can refine the relationships represented in the graph. The fact that Knowledge Graphs are by nature designed to be dynamic means that this should not be a significant issue.

\subsubsection{Entity and Relationship Representation}
The proposed approach to create a granular, interconnected model of the supply chain would involve creating distinct nodes for organizations, products, and services, and encoding additional information within these nodes as attributes. In order to set it up to handle the accumulation and attribution of E-liabilities at different levels, we can define the following for the e-liability knowledge graph:

\begin{itemize}

\item Organization Nodes: Each organization involved in the supply chain would be represented as a node. The attributes for these nodes could include the organization's name, location, industry sector, and any other relevant information. An important attribute would be the organization's total E-liabilities, which can be updated as new information on E-liabilities it inherits comes in.
\item Product/Service Nodes: Each distinct product or service would be a node, with attributes including its name, the organization that produces it, and its individual E-liabilities.
\item Process Nodes: These nodes represent processes within an organization, such as manufacturing or distribution, that generate emissions. These nodes could store information such as the process name, related organization, the emissions associated with this process (which might change based on the product or service being created), and other relevant data.

\end{itemize}
The next representation category to be considered are the edges. As we have described previously, edges represent relationships between nodes. We can identify the following types of edges:

\begin{itemize}
    \item  Production Edges: These would connect an organization node to the product/service nodes it produces. The edges may carry attributes like the quantity of product/service produced and the amount of CO2 emissions per unit of product/service. The latter can be used to update the E-liabilities of the product/service nodes and in turn, the organization node.
    \item Supply Edges: These edges would connect an organization node to the product/service nodes it supplies to other organizations while also being able to carry similar attributes to the production edges.
    \item Process Edges: These would connect an organization node to its internal process nodes or from internal process nodes to product/service. They can carry information on the quantity of product/service each process produces and the associated CO2 emissions. Edges from an organization to a process would indicate that the organization carries out this process. The edges could contain attributes like the quantity of product or service created by this process. Edges from a process to a product/service would indicate that the process creates this product/service, with the edge attributes detailing the associated emissions.
\end{itemize}

\subsubsection{E-liability Attribution in this Computational Framework}
Attribution here refers to how E-liabilities is assigned and inherited across the supply chain and with respect to he E-liability KG, how we can represent this transfer and inheritance. Having defined the entity and relationship aspects of this knowledge graph, we describe the E-liability attribution under this framework as follows:

\begin{itemize}
    \item Product/Service Level: The E-liabilities of each product/service would be the sum of the CO2 emissions associated with its production and supply, as indicated by the attributes of the connecting edges.
    \item Process Level: The E-liabilities of a process would be the sum of the CO2 emissions associated with that process, as indicated by the attributes of the connecting edges.
    \item Organization Level: The E-liabilities of an organization would be the sum of the E-liabilities of all the product/service nodes it produces, or the sum of the E-liabilities of its process nodes, depending on how granular we want the analysis to be. This setup allows us to update the E-liabilities of different entities in the graph as new data comes in. For example, if new data reveals increased CO2 emissions in a certain process, we can update the E-liabilities of the corresponding process node, and then propagate this change up to the organization node and down to the product/service nodes.
\end{itemize}

This approach that we have presented is a generalized one that would need to be tailored to a firm's specific needs and the available data. In order to fully deploy it in the real world, several additional considerations would be necessary. For example, we would need to address data security and privacy issues when dealing with potentially sensitive information on organizations and their operations. There are also other considerations that must be taken into account. We describe some of these next. 

\subsubsection{Additional Considerations for the AI-driven E-Liability Knowledge Graph}
To ensure our E-Liability Knowledge Graph provides the most accurate and granular account of E-liabilities across a supply chain, there are a number of additional considerations and aspects we must address.

\begin{itemize}
\item Handling of Circular Supply Chains: It is important to understand that while some supply chains may be circular in nature, particularly in the context of a circular economy where materials are reused and reintroduced into the system, they still have a linear progression from a temporal and process perspective. Each step in a supply chain, irrespective of material reuse, is distinct and takes place at a unique point in time, generating new emissions with each step. These steps can thus be represented as new process nodes in our E-Liability Knowledge Graph, each with their own accumulated E-liabilities. This detailed tracking and accounting of emissions is fundamental to enable effective management of E-liabilities and ultimately the transition towards net-zero emissions.
To illustrate this, consider a scenario where a car manufacturer, AutoCorp, produces vehicles, then reuses or recycles parts when the vehicle reaches the end of its life. Despite the recycling loop, each step— from receiving raw materials, to production, sale, return, recycling, and reuse— is unique, occurs at a different time, and results in new emissions. As a result, each step forms a distinct process node on the E-Liability Knowledge Graph, with E-liabilities accumulated at each node due to the emissions generated at that step. This level of granularity in tracking allows AutoCorp to pinpoint the emission-intensive steps in their supply chain and strategize targeted carbon reduction or offset efforts. 

\item Balancing Granularity and Complexity: While we seek to create a comprehensive and detailed representation of E-liabilities in a supply chain, we must also be mindful of the complexity this could introduce into the E-Liability Knowledge Graph. It is crucial that we find the right balance between granularity and complexity to ensure the graph remains manageable and practical. We have previously proposed a framework that incorporating organization, product, and process nodes, with processes also forming part of the edge data. This level of granularity may be allow us to gain visibility at the organization and product level while maintaining a manageable complexity. On the other hand, a deeper granularity than this representation may be too difficult to computationally manage.
It is also important that the data used to populate the knowledge graph is reliable and verifiable. We must give careful consideration to the source of this data, and the methods used for its acquisition and verification. This will be discussed further in the next section on sourcing and verifying data.
\end{itemize}

In conclusion, the design of the E-Liability Knowledge Graph must be comprehensive yet practical, allowing us to accurately track the accumulation of E-liabilities at each step in the supply chain, even in the context of circular supply chains and other complexities. However, even with a robust structural framework like the one we have proposed, certain issues in data aquisition and processing may still arise. In the next section, we will look at the challenges and opportunities involved in sourcing data to populate this graph.

\subsection{E-Liability Knowledge Graph Data Acquisition and Processing}

A critical step in constructing an E-Liability Knowledge Graph involves the sourcing, extraction, processing, and integration of multi-dimensional and multi-source data. We have previously discussed the heterogeneous and complex nature of this data.  In order to integrate them, an E-liability Knowledge Graph will have to require the application of sophisticated data acquisition and processing techniques, and advanced computational tools. We discuss how this might occur as follows:

\subsubsection{Data Acquisition}

Data Acquisition is the preliminary stage where various data sources relevant to the E-liability KG framework are identified. This is where AI-based data extraction methods is particularly valuable, especially when applied to alternative, non-traditional data sources. These sources are typically multi-dimensional, ranging from structured data like corporate environmental reports, shipping logs, and operational records, to unstructured data such as news reports, social media posts related to environmental issues, and more. Leveraging such a broad data spectrum ensures a comprehensive understanding of an organization's E-liabilities. For example, data integration from news reports and environmental reports could provide insightful information about an organization's environmental impact\cite{tsalis2020new}. If the E-liability Knowledge Graph is being constructed by the firm for its internal analyses then the data acquisition process is significantly straightforward and directly obtainable in a structured form. However, if knowledge graph is constructed to map out E-liability transfer and inheritance across firms in an industry, e.g. by a policymaker, researcher or independent carbon emissions auditor, then alternative data sources such as shipping logs, news reports are particularly more important. 

\subsubsection{Data Extraction}

Post data acquisition, extraction becomes a crucial process, particularly when handling unstructured data. To this end, advanced AI techniques such as Natural Language Processing (NLP) can be employed. Named Entity Recognition (NER) algorithms, as we have previously introduced, are a subset NLP that can parse entities such as organization names, product identifiers, locations, etc., from textual data.

\subsubsection{Data Processing}
Post extraction, the data should undergo processing to fit into the framework of the Knowledge Graph. Tasks such as data cleaning, resolving inconsistencies, and standardizing formats can be undertaken during this stage. Semantic technologies like RDF (Resource Description Framework) \cite{assi2020data} and OWL (Web Ontology Language) \cite{chen2020review} may also help in facilitating this process as they structure the data into a format that is machine-readable.

\subsubsection{Data Integration}
The final phase of this process involves integrating the refined data into the Knowledge Graph. The nodes and edges of the graph are populated based on the relationships inferred from the processed data. Advanced graph algorithms can be utilized to ensure the graph remains dynamic and accurately mirrors the real-world emissions liability structure. The use of sophisticated computational tools like Apache Hadoop for distributed data storage as Do. et al. describe for their knowledge graph in \cite{do2022building}, Elasticsearch for data indexing, and Apache TinkerPop for creating and querying graph databases can facilitate this entire process, making it scalable and efficient \cite{zamfir2019systems}. Considerations while choosing these tools include scalability, cost, ease of integration with other systems, and support for advanced analytics.

\subsubsection{Computational Tools for Knowledge Graph Management}

The power of computational tools can be harnessed at every stage of the knowledge graph development and management process, ultimately helping to improve efficiency and ensure scalability. Graph databases like Neo4j can be employed for their exceptional capabilities in storing and retrieving complex data relationships in the form of a graph \cite{needham2019graph}. Neo4j has already been used in applications such as healthcare data management with tremendous success - an example being the work described in \cite{tuck2022cancer} about its use in modeling and supporting graph analytics on lung cancer properties. Other data processing tools like Pandas\cite{pandas} and PySpark \cite{pyspark} allow efficient handling of large datasets, facilitating tasks like data cleaning, transformations, and aggregations. Apache TinkerPop, a graph computing framework, is useful in creating and querying the knowledge graph, making the graph data easily navigable and analyzable. For large-scale data storage and processing, distributed systems like Apache Hadoop and Apache Spark prove invaluable, providing scalability, resilience, and enhanced processing capabilities. Additionally, Elasticsearch helps with efficient indexing and search capabilities for quick data retrieval \cite{zamfir2019systems}.

\subsection{Data Verification and Security}

The importance of data accuracy and protecting any potentially sensitivity data cannot be  included in the E-Liability Knowledge Graph cannot under-emphasized. Data elements need to be classified as confidential or otherwise and data verification and and security measures must be put be place after this. We describe additional considerations as follows:

\subsubsection{Data Verification}

Ensuring data accuracy is integral to the reliability of the E-Liability Knowledge Graph. Verification can be undertaken by cross-verification with other data sources, randomly examining data samples if the volumes are large, checksum validations, and by leveraging machine learning models for anomaly detection.

\subsubsection{Data Integrity}

In order to maintain data integrity, the accuracy and consistency of data throughout its lifecycle must be preserved. There are a few techniques which can help ensure the consistency of data and prevent unauthorized data modifications. Some of them include: audit trails of logs, backup and recovery mechanisms, and checksum validations \cite{kumar2019cloud}.

\subsubsection{Data Confidentiality}
Given the sensitivity of certain data, there is also a need to maintain data confidentiality. Data confidentiality measures such as encryption, access control mechanisms, and anonymization techniques can be used protect sensitive data in these knowledge graphs \cite{kumar2019cloud} \cite{yang2020data}. In addition, the General Data Protection Regulation (GDPR) promulgated by the European Union \cite{gdpr} and the California Consumer Privacy Act (CCPA)~\cite{ccpa} are foundational privacy regulations that must be adhered to when dealing with data that may be in these knowledge graphs. Adherence to these and other regional or national privacy regulations is important. 
\\
\\
Thus far, we have described how we might build an E-liability Knowledge Graph, discussing the various elements and additional consideration around data, tools, storage that must be considered for real-world adoption of the framework. We can see that with meticulously acquiring, processing, verifying, and securing data, the E-Liability Knowledge Graph can be a potent tool for providing actionable insights into the attribution of E-liabilities across supply chains and ultimately how they can be managed. With the proposal of this integration pathway already provided, we now highlight limitations and challenges that might affect the real-world implementation of the framework.

\subsection{Potential Limitations and Challenges}

The implementation of the E-Liability Knowledge Graph, despite its potential benefits, does pose several challenges and limitations that should be addressed. These can broadly be divided into issues related to data, computational aspects, and policy considerations.
We summarize three types of challenges, data-related, computational challenges and policy-related. 
\subsubsection{Data-related Challenges}

\begin{itemize}
\item Data Availability and Quality: The effectiveness of E-Liability Knowledge Graph depends on access to reliable and comprehensive data. However, the availability and quality of data related to emissions can vary widely. Even when an organization maintains good internal data, it is still subject to uncertainty or unreliable data on E-liabilities from upstream. Some organizations also might not report their emissions data completely or accurately, due to lack of regulatory mandates or incentives. Additionally, the data may be inconsistent across different sources, creating discrepancies in the knowledge graph.

\item Data Integration: Even when adequate data is available, integrating disparate data sources into a unified graph can be complex. Different data sources might use different formats, standards, or terminologies, requiring significant pre-processing before integration. 

\end{itemize}
\subsubsection{Computational Challenges}

\begin{itemize}

\item Complexity and Scalability: The creation and management of a detailed knowledge graph, especially one that needs to represent a complex system like global emissions liabilities, can be computationally intensive. As the number of entities and relationships in any such graph increases, so does the complexity of managing and querying the graph. This has significant implications in the computational resources required to analyze emissions at industrial, national or global levels. 

\item Security and Privacy: As we have previously mentioned, the E-liability knowledge graph may contain potentially sensitive information about organizations' emissions and related data. Ensuring the security of this data and maintaining privacy where required can be a critical challenge. Appropriate encryption, access control mechanisms, and compliance with data protection regulations will be necessary and at large-scale data volumes, ensuring 24/7 compliance may be challenging.

\end{itemize}

\subsubsection{Policy-related Challenges}

\begin{itemize}
    
\item Regulatory Compliance: Regulations related to emissions reporting vary globally. Ensuring that the E-Liability Knowledge Graph aligns with the varying regulations and standards in different regions will be a challenge.

\item Stakeholder Acceptance: Finally, achieving widespread acceptance and adoption of the knowledge graph among all stakeholders (organizations, regulators, consumers, etc.) might be a challenge. Each stakeholder might have different priorities and concerns that need to be addressed.
\end{itemize}

In spite of these challenges, we believe that the potential benefits of an E-Liability Knowledge Graph makes it a worthwhile pursuit. With a considered approach that addresses these issues, this framework can provide a powerful tool for carbon accounting and the management of carbon emissions. To demonstrate how this might work at a local firm level, we introduce a toy example in the appendix.

\subsection{E-Liability Knowledge Graph Framework: Conclusion and Future Directions}
Concluding this paper, we acknowledge the E-Liability Knowledge Graph as a pioneering computational and data framework for the practical implementation of the E-liabilities accounting method conceptualized by Kaplan and Rammana in \cite{kaplan2021accounting}. This computational and data framework serves the important function of tracking, managing, and evaluating E-liabilities throughout the supply chain. It provides a structure that enables the capturing of environmental impact data at an unprecedented level of detail, offering unparalleled visibility into the ecological footprint of entities across the supply chain. It not only streamlines emissions liability tracking but also fundamentally redefines our capacity to answer key research questions in business and policy-making contexts. In particular, sustainability decision makers can make more informed decisions around the following:

\begin{itemize}
   
 \item \textbf {Transparency in Supply Chains}: The E-Liability Knowledge Graph framework illuminates the hidden environmental costs inherent within supply chains. The framework facilitates insights on how E-liabilities are distributed across different parts of the supply chain. It can identify which suppliers or products have the highest E-liabilities, and how changes in one part of the supply chain might impact the overall E-liabilities. This unprecedented transparency fosters accountability, drives policy intervention, and encourages environmentally conscious business practices.
\item \textbf {Product Life Cycle Analysis}: This comprehensive data structure enables a detailed life cycle analysis of products. We can trace the cumulative environmental impact of a product from raw material extraction to disposal. This level of granularity aids sustainable product design, influencing decisions about material selection, manufacturing techniques, and end-of-life disposal.
\item \textbf{Efficacy of Sustainability Interventions}: The E-Liability Knowledge Graph framework enables us to quantify and evaluate the effectiveness of sustainability initiatives. Due to more rigorous measuring the reduction in E-liabilities, decision makers can assess the real-world impact of these interventions and pinpoint areas for further improvement.
\item \textbf {Policy Impact Analysis}: From a policy perspective, the E-Liability Knowledge Graph framework provides a tool to measure the impact of environmental regulations on E-liabilities. This can serve as an invaluable feedback mechanism for policy measures, guiding future policy formulation.
\item \textbf{Risk Assessment}: Additionally, the framework equips us with a more nuanced understanding of environmental risks associated with specific activities, locations, or business partners. Such risk assessments can inform business strategy and contingency planning.
\end{itemize}

By addressing these crucial research questions, the E-Liability Knowledge Graph framework brings clarity to the complex environmental interactions within supply chains, empowering more informed, responsible, and sustainable decision-making. This novel approach marks a significant advancement in the practical implementation of E-liabilities based carbon accounting, underscoring the transformative potential of data-driven insights for promoting environmental sustainability. By continually refining and adopting such a methodology, we can progress towards a future where carbon emissions are not just properly accounted for, but also actively managed. One key feature of this framework is its ability to incorporate advanced computational techniques such as machine learning and large-volumes of data, which has been a major focus of this study.\\

\bibliographystyle{aaai} \bibliography{aaai23.bib}

\begin{thebibliography}{31}
\providecommand{\natexlab}[1]{#1}

\bibitem[{gdp(2016)}]{gdpr}
 2016.
\newblock Regulation (EU) 2016/679 of the European Parliament and of the Council of 27 April 2016 on the protection of natural persons with regard to the processing of personal data and on the free movement of such data, and repealing Directive 95/46/EC (General Data Protection Regulation).
\newblock Official Journal of the European Union.

\bibitem[{Armbrust et~al.(2015)Armbrust, Xin, Lian et~al.}]{pyspark}
Armbrust, M.; Xin, R.~S.; Lian, C.; et~al. 2015.
\newblock Spark SQL: Relational Data Processing in Spark.
\newblock In \emph{Proceedings of the 2015 ACM SIGMOD International Conference on Management of Data}, 1383--1394. ACM.

\bibitem[{Assi, Mcheick, and Dhifli(2020)}]{assi2020data}
Assi, A.; Mcheick, H.; and Dhifli, W. 2020.
\newblock Data linking over RDF knowledge graphs: A survey.
\newblock \emph{Concurrency and Computation: Practice and Experience}, 32(19): e5746.

\bibitem[{Brander et~al.(2021)Brander, Ascui, Scott, and Tett}]{brander2021carbon}
Brander, M.; Ascui, F.; Scott, V.; and Tett, S. 2021.
\newblock Carbon accounting for negative emissions technologies.
\newblock \emph{Climate Policy}, 21(5): 699--717.

\bibitem[{Brander, Gillenwater, and Ascui(2018)}]{brander2018creative}
Brander, M.; Gillenwater, M.; and Ascui, F. 2018.
\newblock Creative accounting: A critical perspective on the market-based method for reporting purchased electricity (scope 2) emissions.
\newblock \emph{Energy Policy}, 112: 29--33.

\bibitem[{Chen, Jia, and Xiang(2020)}]{chen2020review}
Chen, X.; Jia, S.; and Xiang, Y. 2020.
\newblock A review: Knowledge reasoning over knowledge graph.
\newblock \emph{Expert Systems with Applications}, 141: 112948.

\bibitem[{Do et~al.(2022)Do, Phan, Le, and Gupta}]{do2022building}
Do, P.; Phan, T.; Le, H.; and Gupta, B.~B. 2022.
\newblock Building a knowledge graph by using cross-lingual transfer method and distributed MinIE algorithm on apache spark.
\newblock \emph{Neural Computing and Applications}, 1--17.

\bibitem[{Ducoulombier(2021)}]{ducoulombier2021understanding}
Ducoulombier, F. 2021.
\newblock Understanding the importance of scope 3 emissions and the implications of data limitations.
\newblock \emph{The Journal of Impact and ESG Investing}, 1(4): 63--71.

\bibitem[{Fensel et~al.(2020)Fensel, {\c{S}}im{\c{s}}ek, Angele, Huaman, K{\"a}rle, Panasiuk, Toma, Umbrich, Wahler, Fensel et~al.}]{fensel2020introduction}
Fensel, D.; {\c{S}}im{\c{s}}ek, U.; Angele, K.; Huaman, E.; K{\"a}rle, E.; Panasiuk, O.; Toma, I.; Umbrich, J.; Wahler, A.; Fensel, D.; et~al. 2020.
\newblock Introduction: what is a knowledge graph?
\newblock \emph{Knowledge graphs: Methodology, tools and selected use cases}, 1--10.

\bibitem[{He et~al.(2022)He, Luo, Shamsuddin, and Tang}]{he2022corporate}
He, R.; Luo, L.; Shamsuddin, A.; and Tang, Q. 2022.
\newblock Corporate carbon accounting: a literature review of carbon accounting research from the Kyoto Protocol to the Paris Agreement.
\newblock \emph{Accounting \& Finance}, 62(1): 261--298.

\bibitem[{Heath and Bizer(2011)}]{heath2011linked}
Heath, T.; and Bizer, C. 2011.
\newblock Linked data: Evolving the web into a global data space.
\newblock \emph{Synthesis lectures on the semantic web: theory and technology}, 1(1): 1--136.

\bibitem[{Jiang et~al.(2019)Jiang, An, Gao, Liu, and Xi}]{jiang2019factors}
Jiang, M.; An, H.; Gao, X.; Liu, S.; and Xi, X. 2019.
\newblock Factors driving global carbon emissions: A complex network perspective.
\newblock \emph{Resources, Conservation and Recycling}, 146: 431--440.

\bibitem[{Kaplan and Ramanna(2021)}]{kaplan2021accounting}
Kaplan, R.~S.; and Ramanna, K. 2021.
\newblock Accounting for climate change.
\newblock \emph{Harvard Business Review}, 99(6): 120--131.

\bibitem[{Kaplan, Ramanna, and Roston(2023)}]{kaplan2023accounting}
Kaplan, R.~S.; Ramanna, K.; and Roston, M. 2023.
\newblock Accounting for Carbon Offsets – Establishing the Foundation for Carbon-Trading Markets.
\newblock Working Paper 23-050, Harvard Business School.

\bibitem[{Kumar and Goyal(2019)}]{kumar2019cloud}
Kumar, R.; and Goyal, R. 2019.
\newblock On cloud security requirements, threats, vulnerabilities and countermeasures: A survey.
\newblock \emph{Computer Science Review}, 33: 1--48.

\bibitem[{Legislature(2018)}]{ccpa}
Legislature, C. 2018.
\newblock Assembly Bill No. 375: CHAPTER 55 An act to add Title 1.81.5 (commencing with Section 1798.100) to Part 4 of Division 3 of the Civil Code, relating to privacy. [Approved by Governor June 28, 2018. Filed with Secretary of State June 28, 2018. effective January 1, 2020].
\newblock Available at: \url{https://leginfo.legislature.ca.gov/faces/billTextClient.xhtml?bill_id=201720180AB375}.

\bibitem[{McKinney(2010)}]{pandas}
McKinney, W. 2010.
\newblock Pandas: a Foundational Python Library for Data Analysis and Statistics.
\newblock \emph{Python for High Performance and Scientific Computing}, 14(9): 1--9.

\bibitem[{Mohamed, Nov{\'a}{\v{c}}ek, and Nounu(2020)}]{mohamed2020discovering}
Mohamed, S.~K.; Nov{\'a}{\v{c}}ek, V.; and Nounu, A. 2020.
\newblock Discovering protein drug targets using knowledge graph embeddings.
\newblock \emph{Bioinformatics}, 36(2): 603--610.

\bibitem[{Monyei and Jenkins(2018)}]{MONYEI201848}
Monyei, C.~G.; and Jenkins, K.~E. 2018.
\newblock Electrons have no identity: Setting right misrepresentations in Google and Apple’s clean energy purchasing.
\newblock \emph{Energy Research \& Social Science}, 46: 48--51.

\bibitem[{Needham and Hodler(2019)}]{needham2019graph}
Needham, M.; and Hodler, A.~E. 2019.
\newblock \emph{Graph algorithms: practical examples in Apache Spark and Neo4j}.
\newblock O'Reilly Media.

\bibitem[{Peng et~al.(2023)Peng, Xia, Naseriparsa, and Osborne}]{peng2023knowledge}
Peng, C.; Xia, F.; Naseriparsa, M.; and Osborne, F. 2023.
\newblock Knowledge graphs: Opportunities and challenges.
\newblock \emph{Artificial Intelligence Review}, 1--32.

\bibitem[{Qian et~al.(2017)Qian, Li, Zhang, Chen, Jung, and Han}]{qian2017social}
Qian, J.; Li, X.-Y.; Zhang, C.; Chen, L.; Jung, T.; and Han, J. 2017.
\newblock Social network de-anonymization and privacy inference with knowledge graph model.
\newblock \emph{IEEE Transactions on Dependable and Secure Computing}, 16(4): 679--692.

\bibitem[{Roston et~al.(2022)Roston, Seiger, Seiger, and Heller}]{roston2022road}
Roston, M.; Seiger, A.; Seiger, A.; and Heller, T.~C. 2022.
\newblock The Road to Climate Stability Runs through Emissions Liability Management.
\newblock Available at SSRN or \url{http://dx.doi.org/10.2139/ssrn.4363520}.

\bibitem[{Tsalis et~al.(2020)Tsalis, Malamateniou, Koulouriotis, and Nikolaou}]{tsalis2020new}
Tsalis, T.~A.; Malamateniou, K.~E.; Koulouriotis, D.; and Nikolaou, I.~E. 2020.
\newblock New challenges for corporate sustainability reporting: United Nations' 2030 Agenda for sustainable development and the sustainable development goals.
\newblock \emph{Corporate Social Responsibility and Environmental Management}, 27(4): 1617--1629.

\bibitem[{Tuck(2022)}]{tuck2022cancer}
Tuck, D. 2022.
\newblock A cancer graph: a lung cancer property graph database in Neo4j.
\newblock \emph{BMC Research Notes}, 15(1): 45.

\bibitem[{Wbcsd(2004)}]{wbcsd2004greenhouse}
Wbcsd, W. 2004.
\newblock The greenhouse gas protocol.
\newblock \emph{A corporate accounting and reporting standard, Rev. ed. Washington, DC, Conches-Geneva}.

\bibitem[{Wiedmann and Lenzen(2018)}]{wiedmann2018environmental}
Wiedmann, T.; and Lenzen, M. 2018.
\newblock Environmental and social footprints of international trade.
\newblock \emph{Nature Geoscience}, 11(5): 314--321.

\bibitem[{Williamson and Gattuso(2022)}]{williamson2022carbon}
Williamson, P.; and Gattuso, J.-P. 2022.
\newblock Carbon removal using coastal blue carbon ecosystems is uncertain and unreliable, with questionable climatic cost-effectiveness.
\newblock \emph{Frontiers in Climate}, 4.

\bibitem[{Xu et~al.(2020)Xu, Ruan, Korpeoglu, Kumar, and Achan}]{xu2020product}
Xu, D.; Ruan, C.; Korpeoglu, E.; Kumar, S.; and Achan, K. 2020.
\newblock Product knowledge graph embedding for e-commerce.
\newblock In \emph{Proceedings of the 13th international conference on web search and data mining}, 672--680.

\bibitem[{Yang, Xiong, and Ren(2020)}]{yang2020data}
Yang, P.; Xiong, N.; and Ren, J. 2020.
\newblock Data security and privacy protection for cloud storage: A survey.
\newblock \emph{IEEE Access}, 8: 131723--131740.

\bibitem[{Zamfir et~al.(2019)Zamfir, Carabas, Carabas, and Tapus}]{zamfir2019systems}
Zamfir, V.-A.; Carabas, M.; Carabas, C.; and Tapus, N. 2019.
\newblock Systems monitoring and big data analysis using the elasticsearch system.
\newblock In \emph{2019 22nd International Conference on Control Systems and Computer Science (CSCS)}, 188--193. IEEE.

\end{thebibliography}

\section{Appendix}

\subsection{Toy Example: Tracking E-Liabilities in the AutoFab Supply Chain}

In the context of E-Liability carbon accounting, a detailed mapping of the automobile manufacturing supply chain using a knowledge graph can be a nuanced task, considering the multitude of product nodes, process nodes, organization nodes, and their various relationships. To illustrate how this works, in this section we present a hypothetical case study featuring the operations of a hypothetical car manufacturer, AutoFab, and its associated supply chain from raw material sourcing to final product consumption.

In this case study, we take a deep dive into the supply chain of AutoFab to showcase the granular representation of E-liabilities using the E-Liabilities Knowledge Graph Framework. This comprehensive network encapsulates a diverse set of nodes, each representing different organizations, products, services, and processes. Additionally, the intricate network of relationships between these nodes enables us to track and accumulate E-liabilities at every stage.

\subsubsection{Organization Nodes in the AutoFab Supply Chain}
The Organization Nodes in this supply chain would be as follows:
\begin{itemize}
\item AutoFab: The automobile manufacturer located in Detroit, MI.
\item SteelCo: AutoFab's steel supplier based in Pittsburgh, PA.
\item IronMine Corp: The source of iron ore for SteelCo, situated in Minnesota.
\item RubberInc: The tire supplier of AutoFab, located in Akron, OH.
\item RubberPlantations Ltd: The producer of natural rubber for RubberInc, based in Thailand.
\item EnergyCorp: The electricity provider for AutoFab and SteelCo.
\item AutoDealer LLC: The distributors of AutoFab cars. 
\item Consumer Doreen: Buys an AutoFab 2023 Car 
\end{itemize}

\subsubsection{Defining Inputs, Processes, and Outputs}
1 IronMine Corp
\begin{itemize}
\item Raw material: Iron ore - product node
\item Processes: Mining, Transport to SteelCo
\item Output: Iron supplied to SteelCo
\end{itemize}
2 SteelCo
\begin{itemize}
\item Inputs: Iron from IronMine Corp, Electricity (from the grid)
\item Processes: Smelting and steel production, Transport to AutoFab
\item Output: Steel supplied to AutoFab
\end{itemize}
3 RubberPlantations Ltd
\begin{itemize}
\item Raw material: Rubber - product node
\item Processes: Rubber extraction, Processing, Transport to RubberInc
\item Output: Processed rubber supplied to RubberInc
\end{itemize}
4 RubberInc
\begin{itemize}
\item Inputs: Processed rubber from RubberPlantations Ltd, Electricity (self-generated)
\item Processes: Electricity generation, Tire production, Transport to AutoFab,
\item Output: Tires supplied to AutoFab, Excess electricity (if any)
\end{itemize}
5 AutoFab
\begin{itemize}
\item Inputs: Steel from SteelCo, Tires from RubberInc, Electricity (from the grid)
\item Processes: Automobile assembly, Transport to Dealerships
\item Output: Cars supplied to various dealerships
\end{itemize}
6 AutoDealer LLC
\begin{itemize}
\item Inputs: Cars from AutoFab, Electricity (from the grid)
\item Processes: facilities and operations, vehicle servicing, vehicle transportation
\item Output: Cars sold to consumers
\end{itemize}
7 EnergyCorp
\begin{itemize}
\item Inputs: Electricity
\item Processes: Voltage Transformation, Electricity Operations (Dispatch), Facilities Management.
\item Output:Electricity
\end{itemize}

\subsubsection{Product Nodes}

\begin{itemize}

\item Iron Ore at IronMine Corp: The iron ore when it is at the mining company.
\item Iron Ore at SteelCo: The Iron when it is still with SteelCo.
\item Steel at SteelCo: The Steel at SteelCo.
\item Steel at AutoFab: The steel when it reaches AutoFab, with a distinct node to account for its transition across the supply chain.
\item Electricity at AutoFab: Electricity from the grid purchased via EnergyCorp as received at AutoFab. 
\item Natural Rubber at RubberPlantations Ltd: The harvested rubber when it is still at the plantation.
\item Processed Rubber at RubberPlantations Ltd: The processed rubber after under going rubber production at RubberPlantations Ltd. 
\item Processed Rubber at RubberInc: The processed rubber after being transported to RubberInc.
\item Tires at RubberInc: The tires when they are manufactured by RubberInc.
\item Tires at AutoFab: The tires when they reach AutoFab.
\item Automobile at AutoFab: The finished automobile produced by AutoFab.
\item Automobile at Consumer Doreen: The automobile after being delivered to Consumer Doreen
\end{itemize}

\subsubsection{Building Relationships and Accumulating E-liabilities}
Upon defining the nodes, we can begin to link these entities with edges while accounting for the flow of products, services and associated E-liabilities.

\begin{itemize}

\item  \textbf{IronMine Corp - SteelCo Relationship}

An edge between the `Iron Ore' product node of IronMine Corp and the organization node of SteelCo represents the transaction of iron ore. This edge carries e-liability information, such as the CO2 emissions related to mining and transport. When SteelCo receives the iron ore, a new product node for `Iron Ore' is created under SteelCo, and the E-liability from IronMine Corp's `Iron Ore' node is added to it.

\item  \textbf{SteelCo - AutoFab Relationship}

SteelCo's `Steel' product node carries the cumulative E-liability from its `Iron Ore' node, the process of steel production, and transport. When AutoFab receives the steel, a new product node for `Steel' is created under AutoFab with SteelCo's `Steel' node's E-liability.

\item  \textbf{EnergyCorp - AutoFab Relationship}
A similar process is repeated as with SteelCo to capture AutoFab's electricity usage from the grid. As AutoFab purchases electricity from EnergyCorp, it inherits the E-liabilities associated with the amount of electricity purchased. An edge is used to describe this purchase process with electricity nodes on either end. 

\item  \textbf{EnergyCorp - SteelCo Relationship}
 As SteelCo purchases electricity from EnergyCorp, it also inherits the E-liabilities associated with the amount of electricity purchased. We use an edgeto describe this purchase process with electricity nodes on either end.

 \item  \textbf{EnergyCorp - IronMine Relationship}
 Similarly to SteelCo, as Ironmine purchases electricity from EnergyCorp, it also inherits the E-liabilities associated with the amount of electricity purchased. An edge is also used to describe this purchase process with electricity nodes on either end. 

\item  \textbf{RubberPlantations Ltd - RubberInc - AutoFab Relationships}

RubberInc's `Tires' product node accumulates E-liabilities from the `Processed Rubber' node, tire production, and transport. AutoFab's 'Tires' product node, in turn, inherits these E-liabilities as it receives the `Tires'.

\item \textbf{AutoFab - Dealership - Consumer Relationships}

AutoFab's 'Car' product node holds the cumulative E-liabilities from its 'Steel' and 'Tires' nodes, the assembly process, and transport. When a dealership sells a car to a consumer, the accumulated E-liabilities are passed on to the consumer.
\end{itemize}

\begin{figure*}[h!]
\centering
\includegraphics[width=\linewidth]
{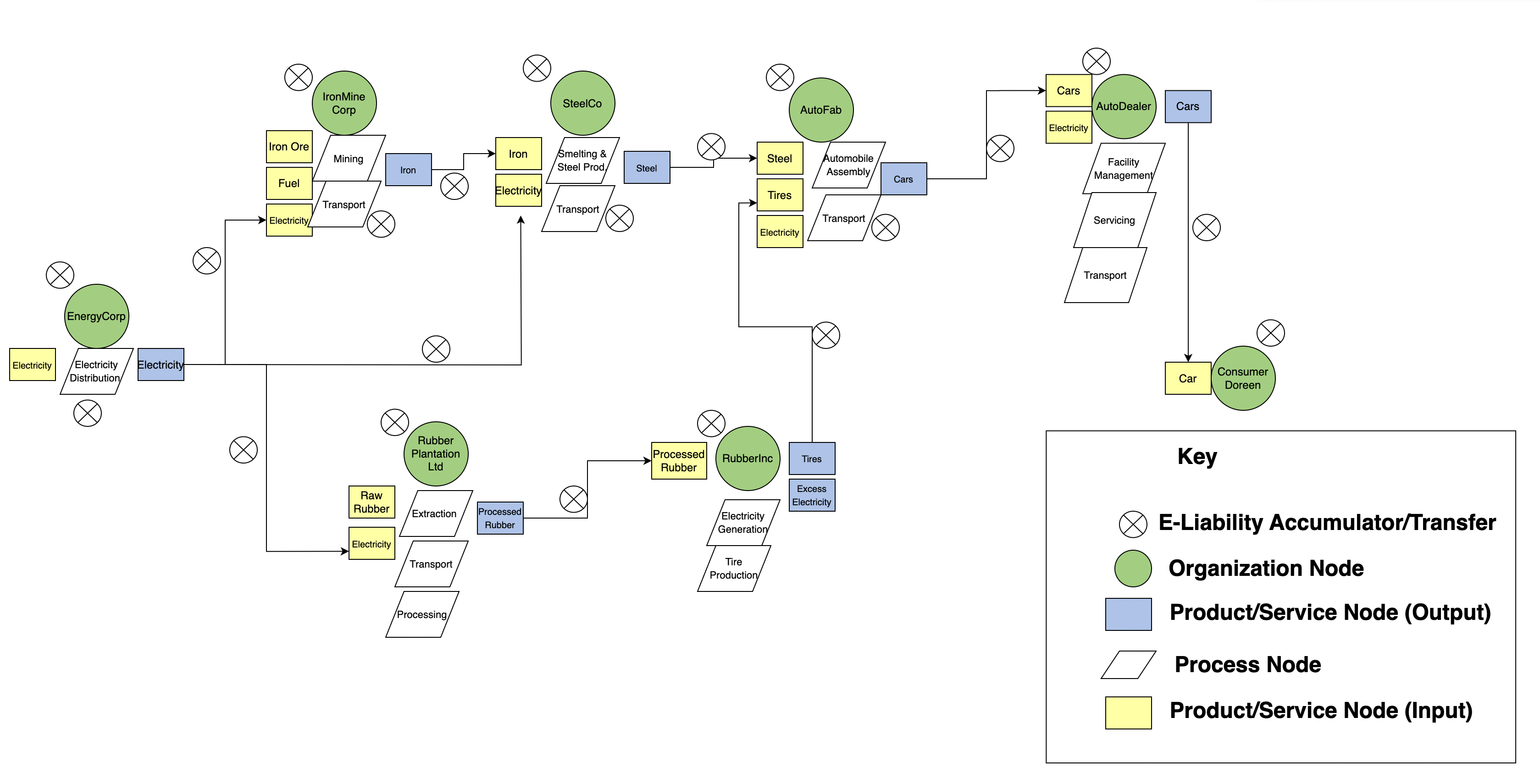}
\caption{Simplified visualization of Knowledge Graph for AutoFab's Supply Chain showing E-liability Accumulation}
\label{autofab}
\end{figure*}

Figure 1 illustrates the various entities and relationships between them in this knowledge graph.

\subsubsection{Tracking E-Liabilities}
This framework enables the precise tracking of E-liabilities throughout the supply chain. For instance, if the process of creating steel at SteelCo and transporting that steel to AutoFab results in an emission of 2 tons of CO2 per ton of steel, and 1 ton of steel is used per car by AutoFab, we can attribute 2 tons of CO2 e-liability to each car inherited due to steel. Similarly, we can also calculate E-liabilities associated with AutoFab's other processes that go into manufacturing its cars such as its electricity usage, and include them in as E-liabilities inherited into the product node of the car produced by AutoFab.  This process continues for the other nodes in the supply chain, each accumulating E-liabilities as materials and products are transferred, processed, and finally sold to the consumer. This resulting in a comprehensive and granular mapping of E-liabilities. It is crucial to note that these are illustrative numbers; the exact quantities would depend on the specific details of the processes involved and the emissions associated with those processes.

By designing our knowledge graph in this manner, we can efficiently track, aggregate, and assign E-liabilities at each step of the supply chain, from raw material extraction to the end consumer.  Ultimately, an E-Liability Knowledge Graph framework allows us to have a visual, structured, and scalable representation of E-liabilities as they move through the supply chain. It provides a granular view of each step of the process, offering insights into which processes or nodes contribute most to the total emissions of a product. By capturing these details, the E-Liability Knowledge Graph framework serves as a tool for understanding and managing the environmental footprint of an organization's supply chain. Businesses can make informed decisions, using the insights gained from this framework, to optimize their operations, reducing their environmental impact and aligning their strategies with sustainability goals. Finally, it is important to recognize that the process of establishing and managing an E-liabilities Knowledge Graph requires rigorous data collection and processing, which can be enhanced with AI/NLP techniques for extracting and interpreting information from diverse data sources.

\end{document}